\documentclass[aps,prl,reprint,superscriptaddress,floatfix]{revtex4-2}

\usepackage{amsmath,amssymb,bm}
\usepackage[T1]{fontenc}
\usepackage{graphicx}
\usepackage{microtype}
\usepackage[colorlinks=true,allcolors=blue]{hyperref}

\begin{document}

\title{Geometric Harmonic Selection of Soft-X-Ray Vortices
from Helical Electron Beams}

\author{Long-Long Zhu}
\affiliation{Department of Physics, Shanghai Normal University,
Shanghai 200234, China}

\author{Longqing Yi}
\affiliation{State Key Laboratory of Dark Matter Physics, Key Laboratory for Laser Plasma (Ministry of Education), Tsung-Dao Lee Institute $\&$ School of Physics and Astronomy, Shanghai Jiao Tong University, Shanghai, China}
\affiliation{ Collaborative Innovation Center of IFSA (CICIFSA), Shanghai Jiao Tong University, Shanghai, China}

\author{Yue-Yue Chen}
\email{yueyuechen@shnu.edu.cn}
\affiliation{Department of Physics, Shanghai Normal University,
Shanghai 200234, China}

\begin{abstract}
We propose a geometric route to quasimonochromatic soft-x-ray vortices through nonlinear Thomson scattering from a finite helical electron bunch. The longitudinal phase difference between successive turns controls their interference, selectively enhancing a matched harmonic while suppressing competing orders. The helical correlation between axial position and azimuth maps the same phase onto an azimuthal winding, linking harmonic selection to orbital angular momentum (OAM). Finite-pulse simulations yield a dominant seventh-harmonic band near $287\,\mathrm{eV}$ with a relative FWHM of $7.1\%$, and its near-axis radiation is predominantly in the $\ell=-7$ mode. The enhanced harmonic order is tunable through pitch matching, while the helix geometry and incident laser wavefront control the vortex charge of the selected harmonic.  This geometric control turns the structured electron source into an active element for shaping the spectrum and wavefront directly at emission.
\end{abstract}

\maketitle

Quasimonochromatic EUV and x-ray vortices combine spectral selectivity with orbital angular momentum (OAM) for resonant spectroscopy \cite{Rouxel2022,Sacchi2026SOLEIL}, coherent imaging \cite{Pancaldi2024,Wang2023Ptychography}, and OAM-sensitive x-ray--matter interactions \cite{Fanciulli2022,VanVeenendaal2007}. Sources based on high-harmonic generation (HHG) and plasma mirrors typically emit multiple harmonics, requiring spectral selection to isolate a quasimonochromatic vortex \cite{Geneaux2016,Wang2019Plasma}. Undulator/free-electron-laser (FEL) schemes can generate coherent narrowband vortices with controlled OAM but require high-brightness relativistic beams and dedicated beamlines \cite{Ribic2017,Hemsing2020}. Recent single-harmonic enhancement has enabled quasimonochromatic EUV generation from HHG without demonstrated OAM control \cite{Zhu2026QMHHG}. 
A compact
source that combines harmonic selection and OAM control directly at
emission is therefore highly desirable.

Nonlinear Thomson scattering (NTS) provides a platform for this purpose. An intense laser acts as a short-period undulator, enabling comparable photon energies at lower electron energies than in magnetic-undulator sources \cite{Sarachik1970,Esarey1993}. However, nonlinear electron motion produces multiple angle-dependent harmonics \cite{Chen1998}, while finite-pulse and electron-beam effects can broaden spectral lines \cite{Terzic2016,Johnson2022} and angular-momentum distributions \cite{Bogdanov2019,Ababekri2024,Zhou2026,Taira2018}.
Laser chirping can narrow individual harmonics by compensating ponderomotive broadening but does not by itself suppress neighboring orders \cite{Terzic2014,Krafft2023}.
Spectral narrowing, harmonic isolation, and OAM-mode purity therefore remain distinct challenges for compact laser-driven vortex sources.

Electron-beam structuring offers an approach to shaping radiation through interference between fields emitted by different electrons \cite{Hemsing2014RMP,Wang2025Microbunching}. Helical microbunching has been proposed and demonstrated for coherent OAM radiation \cite{Hemsing2011,Hemsing2012,Hemsing2013}. In undulator/FEL schemes, the electron structure encodes the wavefront, while microbunching and radiator resonance jointly select the wavelength \cite{Hemsing2012,Sun2023OAM}. More generally, screw-symmetric electron distributions couple longitudinal and azimuthal phases, yielding correlated spectral and total-angular-momentum selection rules \cite{Bogdanov2020}. In Thomson scattering, longitudinal microbunching combined with laser chirping has been proposed to select a nonlinear harmonic and generate a few-cycle soft-x-ray vortex \cite{Schaap2023}. 
These advances motivate a geometric approach to generating quasimonochromatic x-ray vortices with tunable vortex charge.

\begin{figure}[t]
  \centering
  \includegraphics[width=\columnwidth]{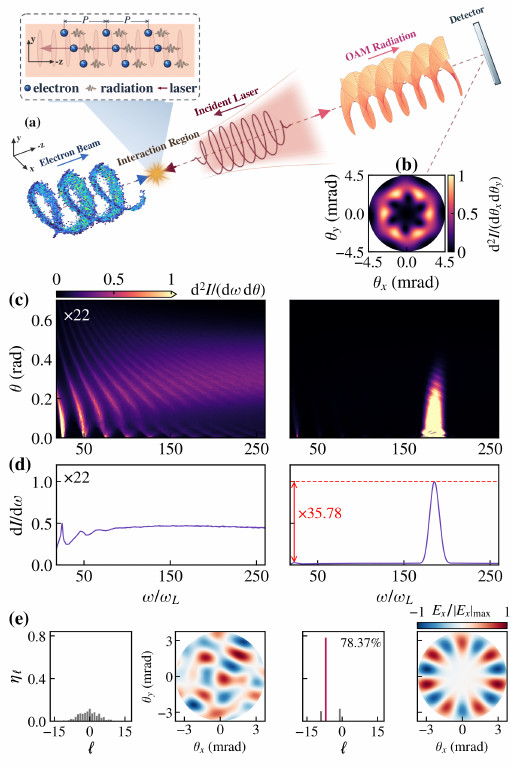}
  \caption{Geometric harmonic selection and OAM control using a helical electron bunch. (a) Head-on collision geometry. The inset illustrates equivalent source elements in adjacent turns, separated by the pitch magnitude $|P|$. (b) H7 angular intensity within $4.5\,\mathrm{mrad}$. (c) Normalized angle-resolved spectra $d^2I/(d\omega\,d\theta)$ versus emission angle $\theta$ (rad) and normalized frequency $\omega/\omega_L$  for the cylindrical (left) and helical (right) bunches. (d) Corresponding angle-integrated spectra over $0\leq\theta\leq0.7\,\mathrm{rad}$. In (c) and (d), the cylindrical-bunch signal is magnified by a factor of 22, with both cases normalized to the helical H7 peak. (e) H7 OAM-mode distributions and corresponding angular distributions of $E_x$ for the cylindrical (left) and helical (right) bunches. Here $\eta_\ell$ denotes the modal power fraction in mode $\ell$, evaluated within $3.75\,\mathrm{mrad}$  and the respective H7 FWHM bands.}
  \label{fig:overview}
\end{figure}

In this Letter, we exploit the axial--azimuthal correlation of a finite helical electron bunch to link harmonic selection and vortex formation in nonlinear Thomson scattering [Fig.~\ref{fig:overview}]. The longitudinal phase that enhances a matched harmonic and suppresses competing orders through interference between successive turns also imprints an azimuthal winding on the emitted field. The pitch therefore determines which harmonic is enhanced and contributes to its vortex charge, which can be further tuned through the incident laser wavefront. Finite-pulse simulations agree with an interference model for the finite helix and demonstrate quasimonochromatic soft-x-ray vortices with tunable charge, establishing the structured electron source as an active phase element for spectral filtering and wavefront synthesis directly at emission.

We consider a circularly polarized laser pulse propagating along
$+z$ and colliding head-on with a helically structured electron
bunch [Fig.~\ref{fig:overview}(a)]. The laser has wavelength
$\lambda_L=800\,\mathrm{nm}$, frequency
$\omega_L=2\pi c/\lambda_L$, waist $w_0=4.0\,\mu\mathrm{m}$,
and normalized amplitude
$a_0=eE_0/(m_ec\omega_L)=3$, with helicity $\epsilon=-1$.
The pulse has a $40T_L$ flat-top with $4T_L$ cosine ramps on
either side, where $T_L=\lambda_L/c$.
The laser pulse interacts with an electron bunch whose initial
density is concentrated around a finite helix.
At each axial position $z$, the transverse density profile is
Gaussian, with its center located at
\begin{equation}
\bm b(\Phi)=
\left(R_0\cos\Phi,\,R_0\sin\Phi,\,\frac{P\Phi}{2\pi}\right),
\label{eq:helix}
\end{equation}
where $\Phi=2\pi z/P$ is the azimuthal parameter.
The helix has radius $R_0$ and $N_h$ turns, with
$0\leq\Phi\leq2\pi N_h$.
The signed pitch $P$ gives the axial displacement per turn,
and its sign determines the handedness.
The transverse Gaussian profiles are weighted by an axial envelope
that vanishes smoothly at both ends of the bunch.
The resulting number density is
\begin{equation}
\begin{aligned}
n_b(x,y,z)={}&
n_0\sin^2\!\left(\frac{\pi z}{N_hP}\right)\\
&\times\exp\!\left[
-\frac{(x-b_x)^2+(y-b_y)^2}{\sigma_\perp^2}
\right],
\end{aligned}
\label{eq:nb}
\end{equation}
for $0\leq z/P\leq N_h$, and zero otherwise.
Here $(b_x,b_y)$ is the transverse center at the chosen $z$,
$\sigma_\perp$ is the radial rms width about that center,
and $n_0$ is the peak density.
The $\sin^2$ factor provides a Hann envelope over the full bunch.
For the reference configuration, we use $R_0=800\,\mathrm{nm}$,
$N_h=21$, $P\simeq-4.32\,\mathrm{nm}$, and
$\sigma_\perp=4\,\mathrm{nm}$, giving an axial length
$L_h=N_h|P|\simeq90.72\,\mathrm{nm}$.
The initial electron positions are sampled independently from this
distribution.
For either helix handedness, all electrons initially propagate along
$-z$ with Lorentz factor $\gamma=8$ and speed $\beta c$,
where $\beta=\sqrt{1-\gamma^{-2}}>0$.
For comparison, we use a nonhelical cylindrical bunch with the same
electron number and initial momenta.
Possible routes toward preparing helical sources include
laser--undulator-induced helical microbunching
\cite{Hemsing2009,Hemsing2012} and laser-driven microplasma
waveguides \cite{Guo2026}.

The electron trajectories are obtained by integrating the relativistic
Lorentz equation in the prescribed laser field. The far-field radiation
is calculated with FaTiDo \cite{Wen2026FaTiDo} by coherently summing
the Liénard--Wiechert fields. 

Our simulation results show that the helical electron structure dramatically reshapes both the spectrum and OAM content of the radiation. Unlike the weak, broadly distributed harmonics from the cylindrical bunch, the helical bunch generates much stronger radiation concentrated in a narrow seventh-harmonic (H7) band [Fig.~\ref{fig:overview}(c)]. Here H$N$ denotes the $N$th scattered harmonic, with $\omega_N$ its
reference on-axis frequency. For the fundamental (H1), the monochromatic
circular-plane-wave approximation gives
$\omega_1\simeq\omega_L[\gamma(1+\beta)]^2/(1+a_0^2)\simeq25.4\,\omega_L$
\cite{Esarey1993,Rykovanov2016}. This spectral selectivity persists after integration over the emission cone ($\theta\leq0.7\,\mathrm{rad}$), with the H7 peak near $185.4\,\omega_L$ ($4.31\,\mathrm{nm}$), an H7/H1 peak-height ratio of $35.78$, and a relative FWHM of $7.1\%$. The H7 band contains approximately $80\%$ of the spectral energy shown in Fig.~\ref{fig:overview}(d).
More importantly, the helical structure also produces a pronounced OAM-mode selectivity. Within a $3.75\,\mathrm{mrad}$ collection cone, the cylindrical bunch exhibits a broad OAM distribution without a dominant mode. In striking contrast, the helical bunch concentrates $78.37\%$ of the H7-band power into the $\ell=-7$ mode, accompanied by a well-defined sevenfold azimuthal field pattern [Fig.~\ref{fig:overview}(e)]. These results demonstrate that the helical geometry enables simultaneous spectral and OAM-mode selection directly at emission.

The H7 enhancement originates from coherent interference between successive turns of the helical electron bunch. Source elements at equivalent positions in neighboring turns have the same transverse coordinates but are separated axially by the signed pitch $P$. Their different laser--electron encounter times and radiation propagation distances give rise to a relative phase
\begin{equation}
\Delta(\omega,\theta)
=\frac{\omega P}{c}\frac{1+\cos\theta}{1+\beta},
\label{eq:phaseadvance}
\end{equation}
where $\theta$ is measured from the initial electron direction.

The interference across the entire helix determines which harmonics are coherently enhanced or suppressed. To describe this effect, we approximate the bunch by its centerline, assuming identical local radiation responses at equivalent positions in different turns. Writing $\Phi=2\pi j+u$, where $j=0,\ldots,N_h-1$ labels the turns and $0\leq u<2\pi$ specifies the azimuth within each turn, the coherent spectral field becomes
\begin{equation}
\begin{aligned}
\mathcal E(\omega,\theta,\phi)
&=\frac{N_e}{2\pi N_h}\int_0^{2\pi}\mathrm{d}u\,
\mathcal K(u)D_H(\Delta,u),\\
D_H(\Delta,u)
&=\sum_{j=0}^{N_h-1}h_j(u)e^{ij\Delta},
\end{aligned}
\label{eq:turn-factor}
\end{equation}
where $h_j(u)=1-\cos[(2\pi j+u)/N_h]$ accounts for the axial density envelope. The kernel $\mathcal K(u)$ describes the single-electron radiation response and phase variation within one turn, while $D_H$ captures the coherent interference among different turns; see \cite{SM} for details.

Constructive interference occurs when $\Delta$ is an integer multiple of $2\pi$ [Eqs.~\eqref{eq:phaseadvance} and \eqref{eq:turn-factor}]. For on-axis H7 emission, the first nonzero matching condition $|\Delta(\omega_7,0)|=2\pi$ gives the pitch magnitude $P_1=\pi c(1+\beta)/\omega_7$. More generally, the matched pitches satisfy $|P|=sP_1$, where $s=1,2,\ldots$ is the matching order. For the reference configuration $P=-P_1$, we have
$\Delta_7\equiv\Delta(\omega_7,0)=-2\pi$. In the ideal equally spaced
harmonic approximation, $\omega_N=N\omega_7/7$, giving
$\Delta_N\equiv\Delta(\omega_N,0)=-2\pi N/7$ for the $N$th harmonic. For $N_h=21$ turns, the interference factor becomes
\begin{equation}
D_H\!\left(-\frac{2\pi N}{7},u\right)
=\begin{cases}
21,&N=7,14,21,\ldots,\\
0,&\text{otherwise}.
\end{cases}
\label{eq:seven-turn}
\end{equation}
Thus, H7 and its integer multiples interfere constructively, while all other harmonics cancel in this ideal model. Their relative strengths depend on the single-electron radiation response $\mathcal K(u)$. At matching, the coherent field scales as $N_e$, yielding an intensity proportional to $N_e^2$.

The finite length of the helix also contributes to the spectral width of the enhanced harmonic. A small frequency detuning $\delta\omega$ from a matched harmonic frequency $\omega_N$ introduces a phase mismatch
$\delta_\varphi\simeq2\pi s\,\operatorname{sgn}(P)\,\delta\omega/\omega_N$
between successive turns, which accumulates along the helix and limits the bandwidth of constructive interference. Defining the dimensionless mismatch $q=N_h\delta_\varphi/(2\pi)$, the $\sin^2$-weighted interference yields the normalized spectral response and relative bandwidth
\begin{equation}
\begin{gathered}
R_H(q)=\left|\frac{\sin(\pi q)}{\pi q(1-q^2)}\right|^2,
\left(\frac{\Delta\omega_{\rm FWHM}}{\omega_N}\right)_{\rm array}
\simeq\frac{1.44}{N_hs}.
\end{gathered}
\label{eq:array-bandwidth}
\end{equation}
The inverse scaling with $N_hs$ shows that increasing the number of turns or the matching order narrows the selected harmonic. For $N_h=21$ and $s=1$, the estimated relative bandwidth is $6.9\%$, consistent with the simulated angle-integrated H7 bandwidth of $7.1\%$.

\begin{figure}[!tbp]
  \centering
  \includegraphics[width=\columnwidth]{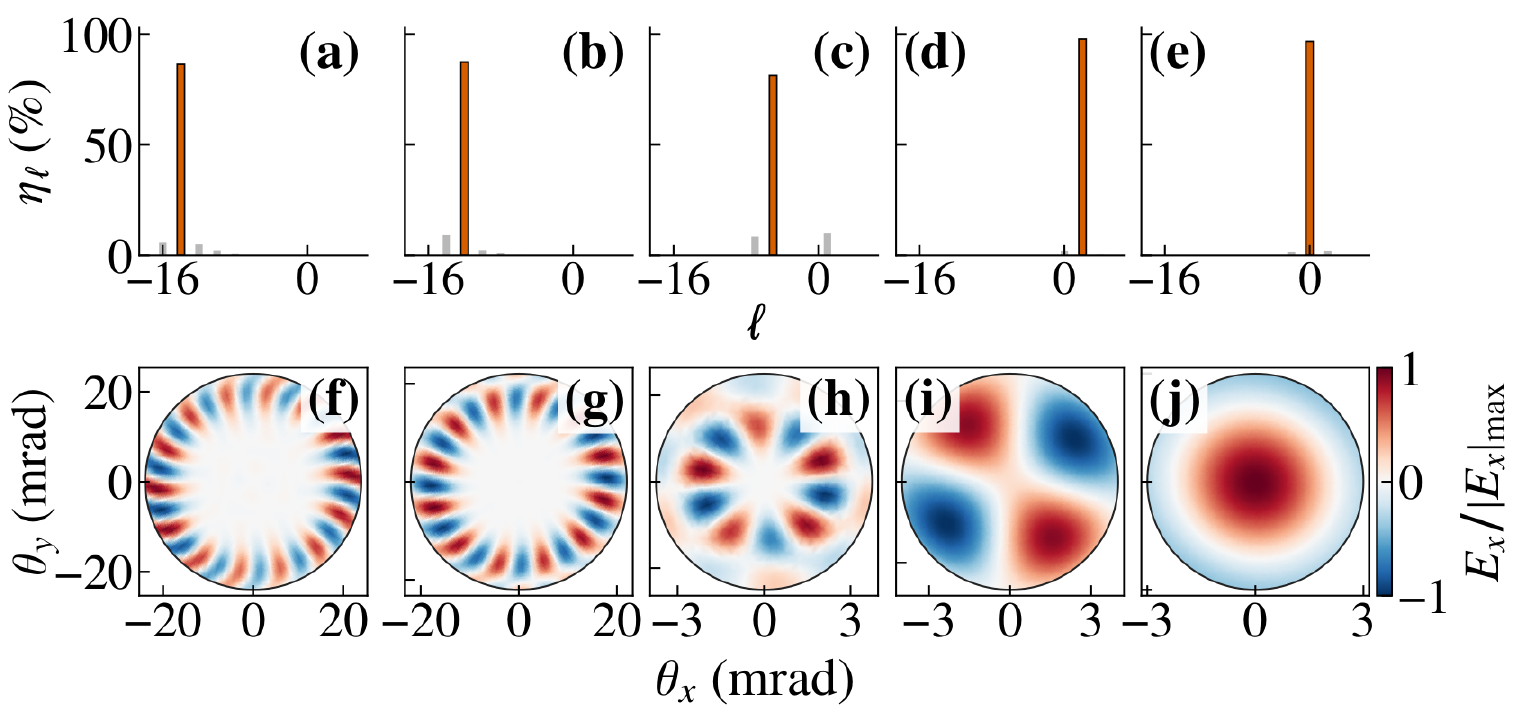}
  \caption{Control of H7 OAM through the incident laser wavefront and signed helix pitch. From left to right, $(m_L,P/P_1)=(-1,-1),(-1,+1),(0,+1),(+1,+1),(+1,-1)$. (a)--(e) H7 OAM-mode distributions, with dominant modes $\ell=-14,-12,-5,+2,0$ and corresponding fractions of $\eta_\ell=86.32\%,87.39\%,81.28\%,97.70\%,96.71\%$. (f)--(j) Corresponding angular distributions of  $E_x$.}
  \label{fig:addition}
\end{figure}

The longitudinal phase responsible for harmonic selection also contributes to the azimuthal structure of the radiation. Along the helix, $z=P\Phi/(2\pi)$, so the phase of the $N$th harmonic varies with azimuth as $\Delta_N\Phi/(2\pi)$. At phase matching,
\begin{equation}
\Delta_N=2\pi s\,\operatorname{sgn}(P),\,
\qquad |P|=s\pi c(1+\beta)/\omega_N,
\label{eq:pitch-root}
\end{equation}
which reduces to $|P|=sP_1$ for H7. The resulting geometric phase $s\,\operatorname{sgn}(P)\Phi$ produces an azimuthal winding whose magnitude is set by $s$ and direction by the sign of $P$. Thus, successive turns can interfere constructively while retaining a nonzero phase winding within each turn.

The vortex charge also receives contributions from the incident laser wavefront and the nonlinear circular-polarization response. An incident laser charge $m_L$ adds a source phase $Nm_L\Phi$, while the circular response contributes an azimuthal phase $(N-1)\epsilon\phi$ in the far field. Combining these contributions gives the topological charge of the selected near-axis circular-polarization component,
\begin{equation}
\ell_N=Nm_L+(N-1)\epsilon
+s\,\operatorname{sgn}(P),
\label{eq:oam-rule}
\end{equation}
where $\epsilon=\pm1$ denotes the laser circular polarization. The derivation is given in the Supplemental Material~\cite{SM}. For the reference H7 configuration ($N=7$), with $m_L=0$, $\epsilon=-1$, $s=1$, and $P<0$, the circular response contributes $-6$ and the helical geometry contributes $-1$, yielding $\ell_7=-7$ [Fig.~\ref{fig:overview}(e)]. Importantly, different matched pitches can preserve constructive interference at the same harmonic while changing its vortex charge, with the incident laser wavefront providing an additional degree of control.

\begin{figure}[!tbp]
  \centering
  \includegraphics[width=\columnwidth]{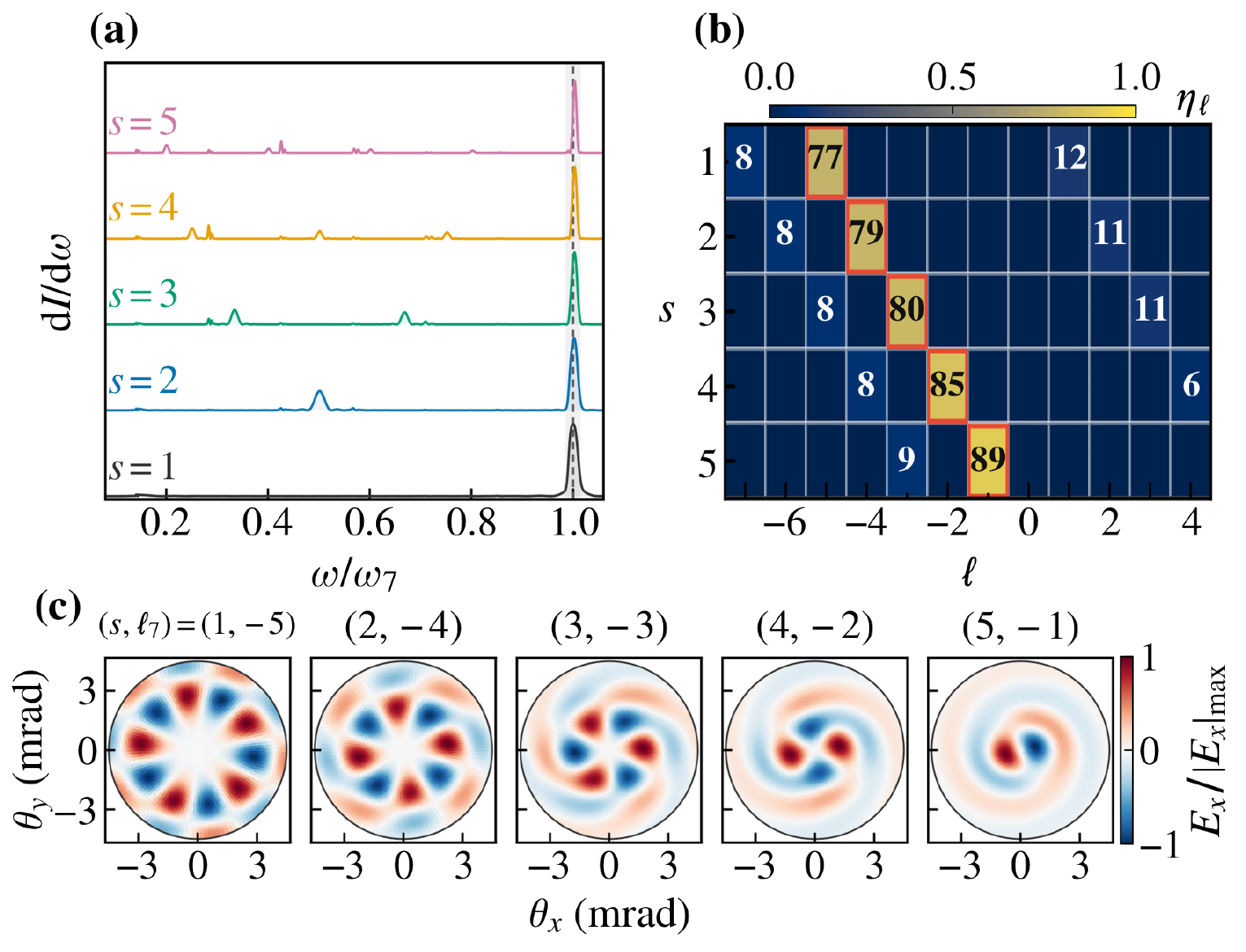}
\caption{Pitch-dependent control of H7 OAM for matched
pitches $P=sP_1$ ($s=1,\ldots,5$), with $m_L=0$,
$\epsilon=-1$, and $\theta\leq4.5\,\mathrm{mrad}$.
(a) Spectra for different $s$, each normalized to its H7 peak.
The dashed line marks $\omega_7$.
(b) H7 OAM-mode distributions. Colors indicate modal
power fractions $\eta_\ell$, while numbers give percentages
($100\eta_\ell$). Outlined cells highlight the dominant
modes $\ell_7=s-6$.
(c) Corresponding angular distributions of 
$E_x/|E_x|_{\max}$ for $s=1,\ldots,5$
(left to right), each normalized independently.}
  \label{fig:roots}
\end{figure}

Figure~\ref{fig:addition} demonstrates control of the H7 vortex charge through the incident laser wavefront and helix handedness at fixed helicity $\epsilon=-1$. Reversing the pitch from $P=-P_1$ to $+P_1$ preserves H7 phase matching but changes the geometric contribution from $-1$ to $+1$, shifting the dominant mode from $\ell=-7$ to $\ell=-5$ for $m_L=0$. In contrast, a unit change in the incident laser charge $m_L$ shifts $\ell_7$ by seven without altering the phase-matching condition. The simulated OAM distributions agree with Eq.~\eqref{eq:oam-rule} [Fig.~\ref{fig:addition}].
The contributions can also cancel. For $(m_L,\epsilon,P/P_1)=(+1,-1,-1)$, Eq.~\eqref{eq:oam-rule} gives $\ell_7=7-6-1=0$. The resulting H7 radiation is centrally peaked, with $96.71\%$ of its modal power in the $\ell=0$ component within $\theta\leq3\,\mathrm{mrad}$ [Figs.~\ref{fig:addition}(e) and \ref{fig:addition}(j)]. The angular distributions also become more concentrated toward the axis as $|\ell|$ decreases, demonstrating control of both the vortex charge and the emission angle within the selected harmonic band.

The matched helix pitch provides a further means of tuning the H7 vortex charge at fixed laser parameters. Figure~\ref{fig:roots} compares five positive-pitch configurations $P=sP_1$ ($s=1,\ldots,5$) with $m_L=0$ and $\epsilon=-1$. All five configurations retain selective enhancement of H7 [Fig.~\ref{fig:roots}(a)], but differ in their geometric phase winding. Equation~\eqref{eq:oam-rule} predicts $\ell_7=s-6$, giving a unit-step change in vortex charge between successive matching orders. The simulations confirm the sequence $\ell=-5,-4,-3,-2,-1$, with dominant modal fractions increasing from approximately $77\%$ to $89\%$ [Fig.~\ref{fig:roots}(b)], accompanied by corresponding changes in the transverse field structure [Fig.~\ref{fig:roots}(c)]. Thus, varying the laser charge $m_L$ shifts $\ell_7$ in steps of seven, whereas changing the matched pitch allows unit-step tuning, providing complementary coarse and fine control of the H7 OAM.

\begin{figure}[b]
  \centering
  \includegraphics[width=\columnwidth]{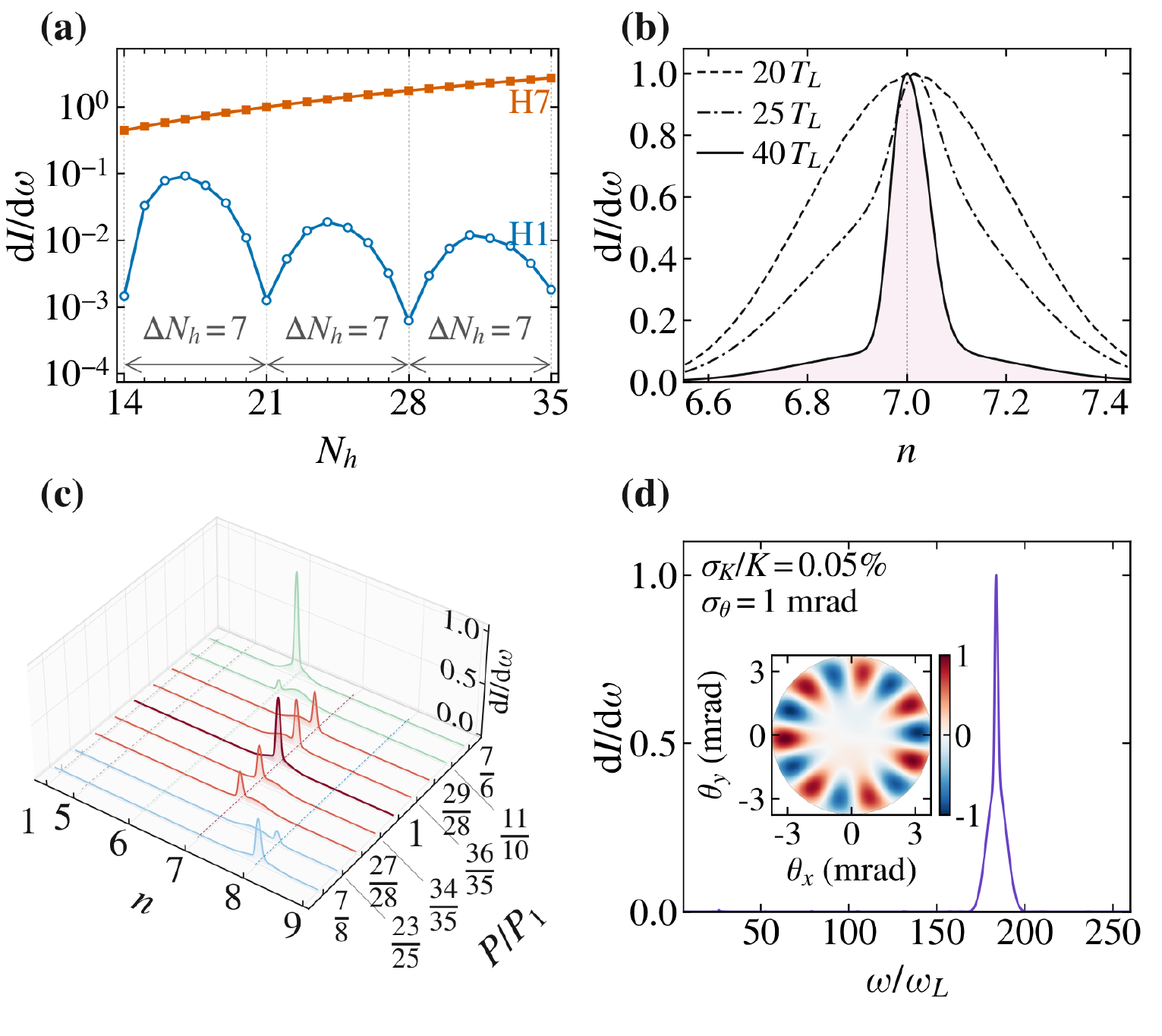}
  \caption{ (a) H1 and H7 peak intensities versus the number of helical turns $N_h$, with $N_e=(3\times10^5)N_h/21$. All intensities are normalized to the H7 peak at $N_h=21$. (b) Normalized H7 spectra for laser durations of $20T_L$, $25T_L$, and $40T_L$. The normalized plotting frequency is $n=7\omega/\omega_7^{\mathrm{ref}}$, where $\omega_7^{\mathrm{ref}}$ is the H7 peak frequency for $40T_L$. (c) Normalized spectra for pitches matched to H6, H7, and H8. The frequency interval $0\leq n\leq5$ is compressed eightfold for clarity. (d) Normalized spectrum with an energy spread of $0.05\%$ and angular spread of $1\,\mathrm{mrad}$. Inset: normalized transverse field within $\theta\leq3.75\,\mathrm{mrad}$.}
  \label{fig:coherence}
\end{figure}

Figure~\ref{fig:coherence} examines the coherence and robustness of the helical source. As the number of turns increases with $N_e\propto N_h$, the H7 peak intensity grows approximately as $N_e^2\propto N_h^2$ [Fig.~\ref{fig:coherence}(a)], consistent with coherent addition of the emitted fields. In contrast, H1 exhibits minima at $N_h=14,21,28$, and $35$, as expected from the weighted cancellation condition, with progressively weaker maxima between them.
The spectral response also depends on the laser duration and helix pitch. Increasing the laser flat-top duration from $20T_L$ to $40T_L$ narrows the H7 peak [Fig.~\ref{fig:coherence}(b)], reflecting the longer coherent emission interval. The pitch scan shows maximum H7 enhancement at $|P|=P_1$, whereas $|P|=7P_1/6$ and $7P_1/8$ favor H6 and H8, respectively [Fig.~\ref{fig:coherence}(c)]. A small pitch deviation $\delta P$ introduces an adjacent-turn phase mismatch $\delta\Delta_7=(\omega_7/c)[(1+\cos\theta)/(1+\beta)]\delta P$, which accumulates along the helix, weakening constructive interference and modifying the spectral side lobes.
Finally, we examine the effect of finite electron-beam spreads. For a relative kinetic-energy spread $\sigma_K/K=0.05\%$ and a total angular spread $\sigma_\theta=1\,\mathrm{mrad}$, the H7 spectral peak remains distinct, while the transverse field retains a sevenfold azimuthal pattern [Fig.~\ref{fig:coherence}(d)].

In conclusion, a finite helical electron bunch converts spatial correlations among emitters into a means of controlling nonlinear radiation. Interference between successive turns selectively enhances and narrows a matched harmonic, while the same geometric phase contributes to its azimuthal winding and OAM. The H7 case demonstrates that different matched pitches can preserve harmonic enhancement while tuning the vortex charge and angular distribution, with the incident laser wavefront providing additional control. Moreover, varying the pitch allows different harmonic orders to be selected. By shaping the spectrum and wavefront directly at emission, this approach offers a route to quasimonochromatic soft-x-ray vortices for OAM-sensitive scattering and structured-illumination imaging. More broadly, it links nonlinear frequency conversion with source-level synthesis of structured high-frequency radiation.

\bibliographystyle{apsrev4-2}
\bibliography{references}

\end{document}


\maketitle
\tableofcontents
\bigskip

This Supplemental Material presents a method for producing the helical electron source considered in the main text and derives the retarded potentials for its collision with a laser, yielding selection rules for finite-coil interference and azimuthal modes. We first describe how an annular electron bunch with a longitudinal Hann envelope acquires a helical density modulation through harmonic coupling in a helical undulator. We then derive the pitch-matching condition and its connection to the vortex phase from classical electron trajectories and the far-field integral.

\section{Generation of helical microbunching in an annular electron beam}
\label{sec:preparation}

In 2009, Hemsing \textit{et al.} developed the theory of harmonic coupling and validated it with three-dimensional simulations~\cite{hemsing2009,hemsing2010erratum}. Helical microbunching of a relativistic electron beam was observed through coherent transition radiation in 2012~\cite{hemsing2012}, and coherent optical vortices were generated and measured in a downstream radiator in 2013~\cite{hemsing2013}. These three studies established the theoretical and experimental basis for the sequence from energy modulation through density microbunching to coherent radiation carrying orbital angular momentum (OAM). A helical undulator drives transverse circular oscillations about the mean electron trajectories, causing electrons at different azimuths to sample the radial gradient of the seed field with different phases. This phase difference is first imprinted on the electron energy and then converted by longitudinal dispersion into density ridges that wind around the beam axis. Those studies, however, used initial electron distributions with filled transverse cross sections. Here we apply the same mechanism to an electron bunch with an initial longitudinal Hann envelope and an annular transverse distribution, obtaining a distribution with helical density ridges.

\subsection{Initial annular distribution, undulator trajectory, and resonance}

During preparation, the electron bunch and the seed laser copropagate along $+z_p$. The bunch has a mean Lorentz factor $\gamma_p$ and a mean axial velocity $\bar{\beta}_z c$. We define the equal-time longitudinal coordinate within the bunch as $\zeta=z_p-\bar{\beta}_zct$ and denote the azimuth by $\varphi$. The longitudinal line density of the annular bunch is first shaped into a Hann envelope, after which helical microbunching is imprinted. Let the initial envelope have length $L_i$ and begin at $\zeta_a$, and take a thin annular transverse distribution:
\begin{equation}
\begin{aligned}
 n_i(r,\varphi,\zeta)&=\lambda_e f_{\mathrm{ann}}(r)\mathcal W_i(\zeta),
 &v&=\frac{\zeta-\zeta_a}{L_i},\\
 \mathcal W_i(\zeta)&=
 \begin{cases}
 1-\cos(2\pi v),&0\leq v\leq1,\\
 0,&\text{otherwise},
 \end{cases}\\
 f_{\mathrm{ann}}(r)&=\mathcal N_r\exp\!\left[-\frac{(r-R_0)^2}{2\sigma_r^2}\right],
 &2\pi\int_0^\infty r\,\mathrm{d}r\,f_{\mathrm{ann}}(r)&=1.
\end{aligned}
\label{eq:1}
\end{equation}
Since $\int\mathcal W_i\,\mathrm{d}\zeta=L_i$, $\lambda_e$ is the mean longitudinal line density. Here $R_0$ is the ring radius, and $\sigma_r$ is approximately the radial rms width of the ring when $\sigma_r\ll R_0$. Transverse shaping at the photocathode into a thin ring can provide such a beam; injector designs and subsequent transport of hollow beams have been studied~\cite{halavanau2020}. Under approximately instantaneous, linear photoemission and envelope-preserving transport, a Hann temporal intensity profile of the photocathode drive laser can establish the required longitudinal line density.

We use the helical-undulator parameter defined by the peak magnetic field, $K=eB_u/(m_ec k_u)$, where $k_u=2\pi/\lambda_u$ and $e>0$. Neglecting higher-order feedback of the seed field on the trajectory, the transverse electron trajectory and velocity are
\begin{subequations}
\label{eq:2}
\begin{align}
 \bm r_\perp(z_p)
 &=r(\cos\varphi,\sin\varphi)
 +r_w\bigl[\bm e_x\sin(k_uz_p)-\sigma_u\bm e_y\cos(k_uz_p)\bigr],
 \label{eq:2a}\\
 \bm\beta_\perp(z_p)
 &=\frac{K}{\gamma_p}\bigl[\bm e_x\cos(k_uz_p)+\sigma_u\bm e_y\sin(k_uz_p)\bigr],
 \label{eq:2b}\\
 r_w&=\frac{K}{\gamma_p\bar{\beta}_zk_u}
 \simeq\frac{K\lambda_u}{2\pi\gamma_p},
 \qquad
 \bar{\beta}_z\simeq1-\frac{1+K^2}{2\gamma_p^2}.
 \label{eq:2c}
\end{align}
\end{subequations}
Here $\sigma_u=\pm1$ specifies the undulator handedness. The transverse oscillation radius $r_w$ and the electron-ring radius $R_0$ describe the single-electron motion and the overall density distribution, respectively. Throughout this section, $K$ follows the convention $\beta_\perp=K/\gamma_p$.

For second-harmonic coupling, we take a linearly polarized Gaussian seed laser copropagating along $+z_p$:
\begin{equation}
 \bm E_s=\bm e_x E_s(r)\cos\Theta,
 \qquad E_s(r)=E_{s0}e^{-r^2/w_{\mathrm{seed}}^2},
 \qquad \Theta=k_sz_p-\omega_st+\psi_0,
 \label{eq:3}
\end{equation}
where $k_s=\omega_s/c=2\pi/\lambda_{\mathrm{seed}}$. In the bunch coordinates, its phase is
\begin{equation}
 \Theta=-k_{\mathrm{slip}}z_p+k_\zeta\zeta+\psi_0,
 \qquad k_{\mathrm{slip}}=k_s\left(\frac{1}{\bar{\beta}_z}-1\right),
 \qquad k_\zeta=\frac{k_s}{\bar{\beta}_z}.
 \label{eq:4}
\end{equation}
The rate of change of the seed phase sampled by the electrons is set by $k_{\mathrm{slip}}$. Because the electrons travel more slowly than light, they slip longitudinally relative to the seed field. If this slippage is synchronized with the $h$th undulator harmonic, $k_{\mathrm{slip}}=hk_u$, then
\begin{equation}
 \lambda_{\mathrm{seed}}
 =\frac{\lambda_u}{h}\left(\frac{1}{\bar{\beta}_z}-1\right)
 \simeq\frac{\lambda_u(1+K^2)}{2h\gamma_p^2}.
 \label{eq:5}
\end{equation}
This is the resonance condition for the preparation stage.

\subsection{Electron-beam azimuthal phase induced by the radial gradient force}

Although an axisymmetric optical field has no intrinsic azimuthal phase, azimuth-dependent terms appear when the electrons sample it along their oscillatory trajectories. For $r_w\ll R_0,w_{\mathrm{seed}}$ and a field that varies slowly along the modulator,
\begin{equation}
\begin{aligned}
 E_s(\bm r_\perp)
 &=E_s(r)+\delta\bm r_w\cdot\bm\nabla_\perp E_s(r)
 +O(r_w^2\nabla_\perp^2E_s)\\
 &=E_s(r)+r_w E_s'(r)\sin(k_uz_p-\sigma_u\varphi)+\cdots.
\end{aligned}
\label{eq:6}
\end{equation}
The energy change arises from the work done by the laser on the electrons. With the electron charge $q_e=-e$,
\begin{equation}
 \frac{\mathrm{d}\gamma_e}{\mathrm{d}z_p}
 =-\frac{e}{m_ec^2\bar{\beta}_z}\bm\beta_\perp\cdot\bm E_s.
 \label{eq:7}
\end{equation}
The term without the gradient accumulates near $h=1$. The gradient term contains both the $k_u$ oscillation of the transverse velocity and the $k_u$ oscillation from trajectory-dependent sampling. Using the identity
\begin{equation}
\begin{aligned}
 &\cos(k_uz_p)\sin(k_uz_p-\sigma_u\varphi)\cos\Theta\\
 &\quad=\frac14\bigl\{
 \sin(2k_uz_p-\sigma_u\varphi+\Theta)
 +\sin(2k_uz_p-\sigma_u\varphi-\Theta)
 -2\sin(\sigma_u\varphi)\cos\Theta\bigr\},
\end{aligned}
\label{eq:8}
\end{equation}
we find that, near the $h=2$ resonance, only the first term has a slowly varying phase, $\Delta_2z_p+k_\zeta\zeta-\sigma_u\varphi+\psi_0$, where $\Delta_2=2k_u-k_{\mathrm{slip}}$. The remaining terms average out over many undulator periods. Integrating over a uniform modulator of length $L_u=N_u\lambda_u$ and retaining the resonant term gives
\begin{equation}
\begin{aligned}
 \Delta\gamma(r,\varphi,\zeta)
 &=-\frac{eKr_wL_u}{4\gamma_p\bar{\beta}_zm_ec^2}
 E_s'(r)\operatorname{sinc}\!\left(\frac{\Delta_2L_u}{2}\right)\\
 &\quad\times\sin\!\left(k_\zeta\zeta-\sigma_u\varphi+\psi_0+\frac{\Delta_2L_u}{2}\right),
 \qquad \operatorname{sinc}x\equiv\frac{\sin x}{x}.
\end{aligned}
\label{eq:9}
\end{equation}
The constant phase and overall sign can therefore be absorbed into $\psi$, allowing the modulation to be written as
\begin{equation}
 \Delta\gamma=A_\gamma(r)\cos\chi_b,
 \qquad\chi_b=k_\zeta\zeta+l_b\varphi+\psi,
 \qquad l_b=-\sigma_u.
 \label{eq:10}
\end{equation}
At resonance and in the ultrarelativistic approximation, the positive modulation amplitude is
\begin{equation}
 A_\gamma(r)\simeq
 \frac{eK^2N_u\lambda_u^2}{8\pi\gamma_p^2m_ec^2}|E_s'(r)|.
 \label{eq:11}
\end{equation}
Equation~\eqref{eq:11} agrees with the gradient-coupling expression used in the 2013 experiment~\cite{hemsing2013}.

More general harmonic coupling can be represented by a gradient operator carrying an azimuthal phase. For the branch corresponding to the chosen undulator handedness, define
\begin{equation}
 \mathcal D_{\sigma_u}
 =e^{-i\sigma_u\varphi}
 \left(\partial_r-\frac{i\sigma_u}{r}\partial_\varphi\right).
 \label{eq:12}
\end{equation}
Each application lowers the azimuthal index of $F(r)e^{il_s\varphi}$ by $\sigma_u$. The leading gradient coupling at the $h$th harmonic involves $h-1$ such operations, giving
\begin{equation}
 l_b=l_s-\sigma_u(h-1).
 \label{eq:13}
\end{equation}
Reversing the handedness selects the opposite branch, yielding the selection rule often written as $l_s\pm(h-1)$ in the literature~\cite{hemsing2009,hemsing2010erratum}. A Gaussian seed with $l_s=0$ at $h=2$ therefore produces a single-strand modulation with $|l_b|=1$.

\subsection{Matching the ring radius to the coupling strength}

The radial gradient of a Gaussian seed vanishes on axis, whereas its strongest-gradient region forms a ring. A hollow electron beam can therefore be better suited than a beam with a filled cross section to microbunching through this mechanism. Specifically,
\begin{equation}
\begin{aligned}
 |E_s'(r)|&=\frac{2E_{s0}r}{w_{\mathrm{seed}}^2}e^{-r^2/w_{\mathrm{seed}}^2},\\
 \frac{\mathrm{d}}{\mathrm{d}r}\left[re^{-r^2/w_{\mathrm{seed}}^2}\right]&=0
 \quad\Longrightarrow\quad R_{\mathrm{opt}}=\frac{w_{\mathrm{seed}}}{\sqrt2}.
\end{aligned}
\label{eq:14}
\end{equation}
If the thin ring is located at $R_0\simeq R_{\mathrm{opt}}$, most electrons simultaneously sample this strong-gradient region. Near this radius, the relative variation of the gradient magnitude is
\begin{equation}
 \frac{|E_s'(R_0+\Delta r)|}{|E_s'(R_0)|}
 =1-2\frac{\Delta r^2}{w_{\mathrm{seed}}^2}
 +O\!\left(\frac{\Delta r^3}{w_{\mathrm{seed}}^3}\right),
 \qquad R_0=R_{\mathrm{opt}}.
 \label{eq:15}
\end{equation}
The first-order radial variation vanishes, so a narrow ring also reduces the spread in modulation amplitudes. Here $w_{\mathrm{seed}}$ is the waist of the seed laser used during preparation.

\subsection{Longitudinal dispersion: From energy modulation to density microbunching}

Energy modulation alone does not form density ridges. Longitudinal dispersion produces energy-dependent longitudinal displacements, causing electrons to accumulate at specific phases of the energy modulation. Let the initial relative energy deviation be $\delta_0$ and the deviation after modulation be $\delta=\delta_0+a_\delta(r)\cos\chi_b$, where $a_\delta=A_\gamma/\gamma_p$. Neglecting transverse--longitudinal coupling and higher-order transport terms,
\begin{equation}
 \zeta_f=\zeta_i+R_{56}\bigl[\delta_0+a_\delta(r)\cos\chi_b\bigr],
 \qquad R_{56}=\frac{\partial\zeta_f}{\partial\delta}.
 \label{eq:16}
\end{equation}
The sign of $R_{56}$ depends on the longitudinal-coordinate convention and the transport system. For a low-energy beam, the undulator or drift itself can provide appreciable dispersion. The 2012 experiment required no additional dispersive section~\cite{hemsing2012}, whereas the 2013 experiment used a magnetic bending section to convert energy modulation into density microbunching~\cite{hemsing2013}.

First consider a cold beam, $\delta_0=0$, and define $C(r)=k_\zeta R_{56}a_\delta(r)$. Particle-number conservation and the Jacobian of Eq.~\eqref{eq:16} give
\begin{equation}
\begin{aligned}
 n_f\,\mathrm{d}\zeta_f&=n_i\,\mathrm{d}\zeta_i,
 \qquad\frac{\partial\zeta_f}{\partial\zeta_i}=1-C(r)\sin\chi_b,\\
 n_f&=\frac{\lambda_e f_{\mathrm{ann}}(r)\mathcal W_i(\zeta_i)}{1-C(r)\sin\chi_b}\\
 &\simeq\lambda_e f_{\mathrm{ann}}(r)\mathcal W_i(\zeta_f)
 \bigl[1+C(r)\sin\chi_{b,f}\bigr],
 \qquad |C|\ll1.
\end{aligned}
\label{eq:17}
\end{equation}
Here $\chi_{b,f}=k_\zeta\zeta_f+l_b\varphi+\psi$. Replacing the coordinate in the phase introduces only a second-order correction. Replacing the envelope additionally requires $k_\zeta L_i\gg1$ and $|R_{56}a_\delta|\ll L_i$, so that its variation over the dispersive displacement can be neglected. The density and energy modulations differ in phase by one quarter of a cycle.

For a finite energy spread, the complex bunching factor gives a more complete relation. Treat the Hann envelope as constant in a local average over one period, and take the initial $\chi_b$ to be uniformly distributed on $[0,2\pi)$. Let $\delta_0$ be an independent Gaussian variable with rms $\sigma_\delta$, and assume $|R_{56}\sigma_\delta|\ll L_i$. Defining $B=k_\zeta R_{56}\sigma_\delta$, we obtain
\begin{equation}
\begin{aligned}
 \mathfrak b_n(r)&\equiv\left\langle e^{-in\chi_{b,f}}\right\rangle\\
 &=\frac{1}{2\pi}\int_0^{2\pi}\mathrm{d}\chi_b\,
 e^{-in\chi_b-inC(r)\cos\chi_b}
 \left\langle e^{-ink_\zeta R_{56}\delta_0}\right\rangle\\
 &=(-i)^nJ_n[nC(r)]\exp\!\left(-\frac{n^2B^2}{2}\right),
 \qquad n\geq1.
\end{aligned}
\label{eq:18}
\end{equation}
The Gaussian average produces the energy-spread suppression factor, while the phase integral yields the ordinary Bessel function. The corresponding density is
\begin{equation}
 n_f(r,\varphi,\zeta_f)\simeq
 \lambda_e f_{\mathrm{ann}}(r)\mathcal W_i(\zeta_f)
 \left[1+2\operatorname{Re}\sum_{n=1}^{\infty}
 \mathfrak b_n(r)e^{in\chi_{b,f}}\right].
 \label{eq:19}
\end{equation}
Thus, the sinusoidal amplitude of the fundamental density modulation is $2|\mathfrak b_1|$. For $|C|,|B|\ll1$, $\mathfrak b_1\simeq-iC/2$, and Eq.~\eqref{eq:19} reduces to Eq.~\eqref{eq:17}.

\subsection{Connection to the nanoscale helical electron tube}

In the slowly varying envelope approximation, the local modulation maxima are given by $\chi_{b,f}=\chi_{\max}+2\pi q$. The ridges on the thin ring are then
\begin{equation}
\begin{aligned}
 \zeta_f(\varphi)&=\frac{\chi_{\max}-\psi-l_b\varphi+2\pi q}{k_\zeta},\\
 \bm r_q(\varphi)&=\bigl(R_0\cos\varphi,R_0\sin\varphi,\zeta_f(\varphi)\bigr),
 \qquad P_{\mathrm{prep}}=-\frac{2\pi l_b}{k_\zeta}.
\end{aligned}
\label{eq:20}
\end{equation}
For $l_b=-\sigma_u$, this gives $P_{\mathrm{prep}}=\sigma_u\bar{\beta}_z\lambda_{\mathrm{seed}}\simeq\sigma_u\lambda_{\mathrm{seed}}$. The sign of the pitch specifies the handedness, and $|P_{\mathrm{prep}}|$ is the axial separation between adjacent complete coils. If $|l_b|>1$, several azimuthal maxima occur in each cross section at a fixed longitudinal position, so the phase of each strand must also be specified. The $n$th density Fourier component generated by the same dispersive map has longitudinal wave number $nk_\zeta$ and azimuthal index $nl_b$, which increase together. The full density is determined by all the components in Eq.~\eqref{eq:19}.

This mechanism imprints the helical correlation, while the preshaped envelope gives the ridges their longitudinal Hann weighting. Linear longitudinal compression shortens both the pitch and the envelope length while preserving the envelope shape. In the weak-microbunching regime, the annular background remains present. In the derivation below, we take the centerline of the density distribution to be the theoretical helix.

\section{Scattering model, coordinates, and reference trajectory in a circularly polarized field}
\label{sec:scattering-model}

After the preparation stage, the electron density is used as the initial source for the scattering calculation. In the coordinates used for the analytical treatment in this section, the scattering laser propagates along $+z$, and the cold electrons initially propagate along $-z$. The observation direction is
\begin{equation}
  \bm n=(\sin\theta\cos\phi,\sin\theta\sin\phi,\cos\theta),
  \qquad u=\pi-\theta,
  \label{eq:21}
\end{equation}
where $u$ is the backscattering angle measured from the initial electron direction and corresponds to the polar angle used to specify the collection aperture in the main text. The near-axis limit is $u\ll1$.

\subsection{Local field amplitude and conditions of validity}
\label{sec:local-field}

Consider a laser beam propagating along $+z$, with a beam waist much larger than the transverse electron excursion. At relativistic intensities, both electric and magnetic forces determine the electron trajectory, generating harmonics and reshaping the angular distribution. The following derivation does not require the laser beam to be an exact plane wave globally; a local plane-wave approximation is used only to solve the single-electron trajectory. Over the transverse extent of each trajectory, the laser envelope and transverse phase are approximated by their values at the electron's initial position. Electrons with different initial positions therefore experience different local plane-wave fields. Let $\chi=ct-z$. The local vector potential is
\begin{equation}
  \bm A=\frac{m_ec}{e}a
  \bigl[\bm e_x\cos\psi+\epsilon\bm e_y\sin\psi\bigr],
  \qquad \psi=k_L\chi-m_L\Phi_0,
  \qquad \epsilon=\pm1.
  \label{eq:22}
\end{equation}
Here $\Phi_0$ labels the azimuth of a source element, and $m_L$ is the integer azimuthal phase index of the scattering drive. The quantity $a=eE_{\mathrm{loc}}/(m_ec\omega_L)$ is the local normalized amplitude of the circularly polarized electric field. Our convention relating it to $A_0$ is
\begin{equation}
  a=\frac{\xi}{\sqrt2},
  \qquad \xi=\frac{eA_0}{m_ec},
  \qquad a(r)\simeq a_0 e^{-r^2/w_0^2}.
  \label{eq:23}
\end{equation}
For a Laguerre--Gaussian (LG) drive, its own radial envelope is used in place of the Gaussian envelope.

The local plane-wave approximation requires the transverse oscillation amplitude $\rho$ to satisfy $\rho|\nabla_\perp\ln a|\ll1$ and $\rho|\nabla_\perp\Phi_L|\ll1$. Neglecting longitudinal variations due to focusing additionally requires the interaction length to be much shorter than the Rayleigh length. Classical trajectories apply when the recoil satisfies $\hbar\omega_N/(\gamma m_ec^2)\ll1$ and the fractional radiative energy loss satisfies $\Delta E_{\mathrm{rad}}/(\gamma m_ec^2)\ll1$. These approximations are used in the analytical derivation below; the finite-pulse simulations in the main text retain the envelope and the spatial variation of the focused field~\cite{sarachik1970,esarey1993}.

\subsection{Light-front invariant and trajectory integration}
\label{sec:reference-trajectory}

In a plane wave, conservation of transverse canonical momentum and light-front momentum allows the trajectory to be parametrized directly by $\chi$. The initial electron Lorentz factor is $\gamma$, and its velocity is $-\beta c\bm e_z$. We define
\begin{equation}
  \eta_0=\gamma-\frac{p_{i,z}}{m_ec}=\gamma(1+\beta),
  \qquad \bm u_e\equiv\frac{\bm p}{m_ec}.
  \label{eq:24}
\end{equation}
For an electron with charge $q_e=-e$, conservation of transverse canonical momentum gives $\bm p_\perp=e\bm A$. Combining this result with the mass-shell relation yields
\begin{equation}
  \bm u_{e,\perp}=a(\cos\psi,\epsilon\sin\psi),
  \qquad u_{e,z}=u_{e,i,z}+\frac{a^2}{2\eta_0},
  \qquad \gamma(\chi)-u_{e,z}=\eta_0.
  \label{eq:25}
\end{equation}
Along the trajectory, $\mathrm d\chi/\mathrm d(ct)=1-\beta_z=\eta_0/\gamma(\chi)$, so that
\begin{equation}
  \frac{\mathrm d\bm r_\perp}{\mathrm d\chi}
  =\frac{\bm u_{e,\perp}}{\eta_0},
  \qquad \frac{\mathrm dz}{\mathrm d\chi}=F_z,
  \qquad F_z=\frac{a^2}{2\eta_0^2}-\frac{\beta}{1+\beta}.
  \label{eq:26}
\end{equation}
Taking the oscillation center as the transverse origin and integrating gives
\begin{equation}
  \begin{aligned}
    \bm r_{0\perp}(\chi)
      &=\rho\bigl[\bm e_x\sin\psi-\epsilon\bm e_y\cos\psi\bigr],
      &\rho&=\frac{a}{\eta_0k_L},\\
    z_0(\chi)&=F_z\chi,
      &ct_0&=\chi+z_0.
  \end{aligned}
  \label{eq:27}
\end{equation}
The integration constants specifying the oscillation center are chosen to be independent of $\Phi_0$.

\section{Far-field integral and translation phase associated with electron--laser encounter times}
\label{sec:far-field-translation}

Coherent summation over different electrons must account for both their optical path differences to the detector and the differences in the times at which they encounter the same laser phase. We use positive frequencies and the Fourier convention
\begin{equation}
  \widetilde{\bm E}(\bm x,k)
  =\frac{1}{\sqrt{2\pi}}\int_{-\infty}^{\infty}
    \mathrm d(ct)\,\bm E(\bm x,t) e^{ ikct},
  \qquad k=\frac{\omega}{c}>0.
  \label{eq:28}
\end{equation}
Integrating the acceleration form of the radiation field by parts gives the velocity form
\begin{equation}
  \begin{aligned}
    \widetilde{\bm E}_{\mathrm{ref}}(\bm n,k)
      &=\frac{ ike}{4\pi\varepsilon_0}
        \frac{ e^{ ikR}}{R\sqrt{2\pi}}
        \int\mathrm d(ct)\,
        \mathsf T_{\bm n}\bm\beta(t)
         e^{ ik[ct-\bm n\cdot\bm r_0(t)]},\\
    \mathsf T_{\bm n}\bm v
      &\equiv\bm n\times(\bm n\times\bm v).
  \end{aligned}
  \label{eq:29}
\end{equation}
The electron charge $q_e=-e$ has been substituted in this expression. If the effective source size is $D$, the far-field expansion requires $R\gg D$ and $kD^2/(2R)\ll1$.

\subsection{Spacetime translation from an equal-time position difference}
\label{sec:equal-time-translation}

Let $\bm b$ be the displacement of an electron relative to a reference electron at the same laboratory time in the incoming free-motion region. If the two electron worldlines, evaluated at the same laser phase $\chi$, differ only by a constant spacetime translation $\Delta^\mu=(\Delta ct,\Delta\bm r)$, preserving $ct-z$ requires $\Delta ct=\Delta z$. In the free-motion region, the equal-time displacement and the spacetime translation also satisfy $\Delta\bm r=\bm b+\bm\beta_i\Delta ct$. Hence
\begin{equation}
  \Delta ct=\frac{b_z}{1-\beta_{i,z}},
  \qquad \Delta\bm r=\bm b+\bm\beta_i\frac{b_z}{1-\beta_{i,z}},
  \qquad \bm\beta_i=-\beta\bm e_z.
  \label{eq:30}
\end{equation}
Substitution into the Fourier kernel of Eq.~\eqref{eq:29} gives the total translation phase
\begin{equation}
  \begin{aligned}
    k(\Delta ct-\bm n\cdot\Delta\bm r)
      &=-k\bm n\cdot\bm b
        +k(1-\bm n\cdot\bm\beta_i)\frac{b_z}{1-\beta_{i,z}}\\
      &=-k\bm n_\perp\cdot\bm b_\perp+kQ(\theta)b_z,
      \qquad Q(\theta)=\frac{1-\cos\theta}{1+\beta}.
  \end{aligned}
  \label{eq:31}
\end{equation}
The two terms in the first line represent the optical-path and encounter-time contributions, respectively; the second line combines them.

\subsection{Phase increment between adjacent coils}
\label{sec:coil-phase-increment}

Electrons at the same azimuth in two adjacent coils have identical transverse coordinates, while their equal-time longitudinal positions differ by one signed axial pitch $P$. Their phase increment at frequency $\omega$ and backscattering angle $u$ is therefore
\begin{equation}
  \Delta(\omega,u)=\frac{\omega P}{c}\frac{1+\cos u}{1+\beta}.
  \label{eq:32}
\end{equation}
This is the intercoil phase relation used in the main text.

\section{Complete harmonic decomposition for a single electron in a circularly polarized field}
\label{sec:single-electron-harmonics}

The single-electron radiation phase contains a mean longitudinal contribution and a transverse oscillatory contribution. The former determines the harmonic centers, while the latter determines the Bessel weights and azimuthal phases of the harmonics. Substituting Eq.~\eqref{eq:27} into Eq.~\eqref{eq:29} gives
\begin{equation}
  ct_0-\bm n\cdot\bm r_0
  =\alpha(\theta)\chi-\rho\sin\theta\sin(\psi-\epsilon\phi),
  \qquad \alpha(\theta)=1+(1-\cos\theta)F_z.
  \label{eq:33}
\end{equation}
The change of variable also gives $\bm\beta\,\mathrm d(ct)=(\mathrm d\bm r_0/\mathrm d\chi)\mathrm d\chi$, so the effective velocity in the integral is
\begin{equation}
  \bm V(\chi)=\frac{a}{\eta_0}
  (\bm e_x\cos\psi+\epsilon\bm e_y\sin\psi)+F_z\bm e_z.
  \label{eq:34}
\end{equation}
Define the fixed circular-polarization basis $\bm e_\epsilon=(\bm e_x+ i\epsilon\bm e_y)/\sqrt2$. The transverse part can be decomposed into terms proportional to $\bm e_\epsilon e^{- i\psi}$ and $\bm e_{-\epsilon} e^{ i\psi}$. The phase exponential is expanded using the Jacobi--Anger identity,
\begin{equation}
  \exp[- ix\sin(\psi-\epsilon\phi)]
  =\sum_{m\in\mathbb Z}J_m(x)
     e^{- im\psi} e^{ im\epsilon\phi},
  \qquad x=k\rho\sin\theta.
  \label{eq:35}
\end{equation}
Multiplying these Fourier components and integrating over $\chi$ yields $2\pi\delta[\alpha k-(m+1)k_L]$, $2\pi\delta[\alpha k-(m-1)k_L]$, and $2\pi\delta(\alpha k-mk_L)$ for the three contributions, respectively. Denoting the positive harmonic index in all three cases by $N$, we obtain
\begin{equation}
  \begin{aligned}
    k_N(\theta)&=\frac{Nk_L}{\alpha(\theta)}
      =\frac{Nk_L(1+\beta)}{D(\theta)},\\
    D(\theta)&=1+\beta\cos\theta
      +(1+\beta)(1-\cos\theta)\frac{a^2}{2\eta_0^2}.
  \end{aligned}
  \label{eq:36}
\end{equation}
Using $\delta[\alpha(k-k_N)]=\delta(k-k_N)/\alpha$, the field can be written as $\widetilde{\bm E}_{\mathrm{ref}}=\sum_{N\ge1}\bm H_N\delta(k-k_N)$, where
\begin{equation}
  \begin{aligned}
    \bm H_N(\Phi_0)
      ={}&\mathcal C_N e^{ iNm_L\Phi_0}\mathsf T_{\bm n}
      \Bigg\{\frac{a}{\sqrt2\eta_0}
      \Big[J_{N-1}(x_N) e^{ i(N-1)\epsilon\phi}\bm e_\epsilon\\
      &\hspace{37mm}
      +J_{N+1}(x_N) e^{ i(N+1)\epsilon\phi}\bm e_{-\epsilon}\Big]
      +F_zJ_N(x_N) e^{ iN\epsilon\phi}\bm e_z\Bigg\},
  \end{aligned}
  \label{eq:37}
\end{equation}
\begin{equation}
  \mathcal C_N=\frac{ ik_Ne}{4\pi\varepsilon_0}
  \frac{ e^{ ik_NR}}{R}\frac{\sqrt{2\pi}}{\alpha},
  \qquad x_N=k_N\rho\sin\theta.
  \label{eq:38}
\end{equation}
Equation~\eqref{eq:37} retains both transverse branches and the longitudinal projection and applies at general observation angles for a local circularly polarized plane wave.

On the backscattering axis, $\alpha(\pi)=(1+a^2)/\eta_0^2$, and hence
\begin{equation}
  k_N(\pi)=\frac{Nk_L\eta_0^2}{1+a^2}
  \simeq\frac{4\gamma^2Nk_L}{1+a^2}
  =\frac{4\gamma^2Nk_L}{1+\xi^2/2}.
  \label{eq:39}
\end{equation}
The last two forms use the ultrarelativistic limit. The strong-field denominator retains the redshift caused by the circularly polarized oscillation. For the dominant near-axis transverse component, $\bm e_\epsilon^*\cdot\mathsf T_{\bm n}\bm e_\epsilon=-1+\sin^2\theta/2$, so its leading azimuthal structure is
\begin{equation}
  H_{N,\epsilon}(\Phi_0)
  \simeq\mathcal G_NJ_{N-1}(x_N)
     e^{ iNm_L\Phi_0}
     e^{ i(N-1)\epsilon\phi}.
  \label{eq:40}
\end{equation}
The factor $\mathcal G_N$ contains the projection and propagation factors. This step neglects projection contributions of higher order in the near-axis expansion.

\section{Coherent angular spectrum of a finite helical source}
\label{sec:finite-helical-source}

The helical source maps the same longitudinal phase onto both intercoil interference and the azimuthal phase within each coil. Let its centerline be
\begin{equation}
  \bm b(\Phi_0)=\left(R_0\cos\Phi_0,R_0\sin\Phi_0,
    \frac{P\Phi_0}{2\pi}\right),
  \qquad 0\le\Phi_0<2\pi N_h.
  \label{eq:41}
\end{equation}
Here $P$ is the signed axial pitch, $N_h$ is the integer number of coils, and the axial length is $L_h=N_h|P|$. The electron number along the centerline is assigned a Hann weight,
\begin{equation}
  \frac{\mathrm dN_e}{\mathrm d\Phi_0}=n_\Phi w_{\mathrm H}(\Phi_0),
  \qquad w_{\mathrm H}(\Phi_0)=1-\cos\frac{\Phi_0}{N_h},
  \qquad n_\Phi=\frac{N_e}{2\pi N_h}.
  \label{eq:42}
\end{equation}
Since $v=\Phi_0/(2\pi N_h)=z/(N_hP)$, this weight corresponds to an axial Hann density profile. The quantity $n_\Phi$ is the mean electron number per unit azimuthal angle, and the total electron number remains $N_e$. The local line density per unit arc length is $\lambda_\ell(\Phi_0)=n_\Phi w_{\mathrm H}(\Phi_0)/\sqrt{R_0^2+(P/2\pi)^2}$. The weight enters the coherent field integral directly. We take $N_h\ge2$ below.

Combining Eq.~\eqref{eq:31} with the incident azimuthal phase gives the phase within a coil,
\begin{equation}
  \Psi_N=-K_N\cos(\Phi_0-\phi)+B_N\Phi_0,
  \qquad K_N=k_NR_0\sin\theta,
  \qquad B_N=Nm_L+\frac{k_NQP}{2\pi},
  \label{eq:43}
\end{equation}
where
\begin{equation}
  k_NQ=N\kappa(\theta),
  \qquad \kappa(\theta)=\frac{k_L(1-\cos\theta)}{D(\theta)}.
  \label{eq:44}
\end{equation}
The quantity $\kappa$ incorporates both the single-electron harmonic frequency and the translation phase between electrons. These two contributions were derived separately in Eqs.~\eqref{eq:36} and~\eqref{eq:31}.

Expanding the transverse propagation phase gives
\begin{equation}
   e^{- iK_N\cos(\Phi_0-\phi)}
  =\sum_{p\in\mathbb Z}(- i)^pJ_p(K_N)
     e^{- ip\Phi_0} e^{ ip\phi},
  \label{eq:45}
\end{equation}
where $p$ is the integer index selected by the source-phase projection. Substitution into Eq.~\eqref{eq:40} and integration with the weight in Eq.~\eqref{eq:42} yield
\begin{equation}
  E^{\mathrm{helix}}_{N,\epsilon}
  =N_e\mathcal G_NJ_{N-1}(x_N)
    \sum_{p\in\mathbb Z}(- i)^pJ_p(K_N)
    \mathcal H_{N_h}(B_N-p)
     e^{ i[p+(N-1)\epsilon]\phi},
  \label{eq:46}
\end{equation}
\begin{equation}
  \mathcal H_{N_h}(y)
  =\frac{1}{2\pi N_h}\int_0^{2\pi N_h}\mathrm d\Phi_0\,
    w_{\mathrm H}(\Phi_0) e^{ iy\Phi_0}
  = e^{ i\pi q}\frac{\sin(\pi q)}{\pi q(1-q^2)},
  \qquad q=N_hy.
  \label{eq:47}
\end{equation}
At $q=0,\pm1$, the continuous limits of the integral definition are used. The Hann kernel $\mathcal H_{N_h}$ determines the matching between the continuous slope $B_N$ and the integer mode $p$, while $J_p(K_N)$ determines the transverse diffraction. For integer $N_h\ge2$, when $B_N=L\in\mathbb Z$, one has $\mathcal H_{N_h}(L-p)=\delta_{pL}$, so only $p=L$ remains. If $B_N=L+\delta$ and the principal-mode amplitude is nonzero, leakage into neighboring modes is given by
\begin{equation}
  \left|\frac{C_{L+r}}{C_L}\right|
  =\left|\frac{J_{L+r}(K_N)}{J_L(K_N)}\right|
    \left|\frac{\delta}{\delta-r}\right|
    \left|\frac{1-N_h^2\delta^2}{1-N_h^2(\delta-r)^2}\right|,
  \qquad r\in\mathbb Z\setminus\{0\}.
  \label{eq:48}
\end{equation}
All removable singularities are evaluated using the corresponding limits of $\mathcal H_{N_h}$.

\subsection{Finite cross section repeated from coil to coil}
\label{sec:repeated-cross-section}

Suppose the cross section around the centerline is described by offsets $\delta\bm b_q(\Phi)$ and normalized weights $w_q$, with $\delta\bm b_q(\Phi+2\pi)=\delta\bm b_q(\Phi)$. The repeated cross section can then be incorporated into a single-coil kernel. Writing $\Phi_0=2\pi j+\Phi$, with $0\le\Phi<2\pi$, the general field is
\begin{equation}
  \begin{aligned}
    \bm E_N&=\sum_{j=0}^{N_h-1}
       e^{ ijN\kappa P}\bm F_{N,j},\\
    \bm F_{N,j}&=\sum_qw_q\int_0^{2\pi}\mathrm d\Phi\,
      w_{\mathrm H}(2\pi j+\Phi)\bm{\mathcal K}_{Nq,j}(\Phi).
  \end{aligned}
  \label{eq:49}
\end{equation}
The single-coil kernel includes the trajectory, local field, propagation phase across the cross section, polarization, and mean line density $n_\Phi$. If these kernels are identical from coil to coil, defining $\bm{\mathcal K}_N=\sum_qw_q\bm{\mathcal K}_{Nq}$ gives
\begin{equation}
  \begin{aligned}
    \bm E_N&=\int_0^{2\pi}\mathrm d\Phi\,
      \bm{\mathcal K}_N(\Phi)D_N^{\mathrm H}(P,\Phi),\\
    D_N^{\mathrm H}(P,\Phi)&=\sum_{j=0}^{N_h-1}
      w_{\mathrm H}(2\pi j+\Phi) e^{ ijN\kappa P}.
  \end{aligned}
  \label{eq:50}
\end{equation}
At exact matching, $N\kappa P\in2\pi\mathbb Z$, and the sum of the coil weights is identically $N_h$.

\section{Pitch branches, the primitive-root condition, and finite bandwidth}

Let $a_N=N\kappa P$ and introduce the auxiliary geometric sum
$D_{N_h}^{(0)}(a)=\sum_{j=0}^{N_h-1}e^{ija}$. Then
\begin{equation}
\begin{aligned}
D_N^{\mathrm H}(P,\Phi)
&=D_{N_h}^{(0)}(a_N)
-\frac{1}{2}e^{i\Phi/N_h}D_{N_h}^{(0)}\left(a_N+\frac{2\pi}{N_h}\right)
\\
&\quad-\frac{1}{2}e^{-i\Phi/N_h}D_{N_h}^{(0)}\left(a_N-\frac{2\pi}{N_h}\right),
\\
D_{N_h}^{(0)}(a)
&=e^{i(N_h-1)a/2}\frac{\sin(N_h a/2)}{\sin(a/2)}.
\end{aligned}
\label{eq:51}
\end{equation}
The continuous limit is taken wherever the denominator vanishes. For the target harmonic order $N_\star$ and the selected observation direction, take
\begin{equation}
P_s=\frac{2\pi s}{N_\star\kappa_\star},
\qquad s\in\mathbb Z\setminus\{0\},
\qquad N_h=MN_\star,
\qquad M=2,3,\ldots.
\label{eq:52}
\end{equation}
At fixed observation direction, field strength, and electron energy, the monochromatic harmonic centers satisfy $k_NQ=N\kappa_\star$. The three geometric sums in Eq.~\eqref{eq:51} give
\begin{equation}
D_N^{\mathrm H}(P_s,\Phi)=
\begin{cases}
N_h, & N_\star\mid Ns,\\
0, & N_\star\nmid Ns.
\end{cases}
\label{eq:53}
\end{equation}
If $d=\gcd(|s|,N_\star)$, the transmitted orders remain multiples of $N_\star/d$. Under the conditions in Eq.~\eqref{eq:52}, the necessary and sufficient condition for complete cancellation of all orders $1\leq N<N_\star$ is again
\begin{equation}
\gcd(|s|,N_\star)=1.
\label{eq:54}
\end{equation}
The condition $M\geq2$ prevents the two frequency-shifted Hann terms from adding constructively at unmatched harmonics. For $N_\star=7$, $N_h=21$, and $s=-1$, we have $M=3$: the ideal weighted intercoil factors vanish for H1--H6, whereas H7 adds constructively. Higher multiples, such as H14 and H21, also add constructively, so the same harmonic family is selected.

On the backscattering axis, Eq.~\eqref{eq:39} gives
\begin{equation}
P_s=\frac{s\lambda_L(1+a^2)}{2N_\star\gamma^2(1+\beta)}.
\label{eq:55}
\end{equation}
For the reference parameters $\gamma=8$ and $R_0/w_0=0.2$, using the on-axis field amplitude of $3.0$ employed in the simulations and a Gaussian focal profile gives $a(R_0)\simeq2.88$. Equation~\eqref{eq:55} then gives $|P_{-1}|\simeq4.2\,\mathrm{nm}$. The simulations in the main text use $|P_{-1}|\simeq4.32\,\mathrm{nm}$ because the single-electron H7 emission is off axis.

\subsection{Detuning, bandwidth, and scans over coil number}

Let $P=P_s+\delta P$ and $\kappa=\kappa_\star+\delta\kappa$. The normalized longitudinal response is
\begin{equation}
\begin{aligned}
\delta\varphi&=N_\star(\kappa_\star\delta P+P_s\delta\kappa),
\\
\left|\mathcal H_{N_h}\left(\frac{\delta\varphi}{2\pi}\right)\right|
&=\left|\frac{\sin(\pi q)}{\pi q(1-q^2)}\right|,
\qquad q=\frac{N_h\delta\varphi}{2\pi}.
\end{aligned}
\label{eq:56}
\end{equation}
The second-order product has been neglected in $\delta\varphi$. The full width at half maximum of the Hann intensity response corresponds to $\Delta q\simeq1.4406$, giving
\begin{equation}
\left(\frac{\Delta\omega_{\mathrm{FWHM}}}{\omega}\right)_{\mathrm{array}}
\simeq\frac{1.44}{N_h|s|}.
\label{eq:57}
\end{equation}
For $N_h=21$ and $|s|=1$, this estimate is $6.9\%$. Coherent addition still requires $N_h|\delta\varphi|$ to be no larger than order unity. For a pitch error alone, the absolute tolerance remains of order $1/(N_hN_\star\kappa_\star)$, while the relative tolerance decreases as $1/|s|$. Across a finite aperture, $\delta\kappa$ arises from variations in the observation direction.

For the first H7 pitch branch and $N_h=7M+\nu$, with $0\leq\nu<7$,
\begin{equation}
D_7^{\mathrm H}=N_h,
\qquad D_1^{\mathrm H}(P_s,\Phi)=0
\qquad (\nu=0,\ M\geq2).
\label{eq:58}
\end{equation}
When the coil number is not an integer multiple of seven, the H1 residual must be calculated from the weighted sum in Eq.~\eqref{eq:51}. If the total electron number is held fixed, $n_\Phi\propto1/N_h$, and the matched H7 field amplitude does not increase with the number of coils. For any fixed field projection, denote the corresponding single-coil kernel by $\mathcal K_N$. Then
\begin{equation}
\left|\frac{\mathcal E_1}{\mathcal E_7}\right|^2
=\frac{
\left|\displaystyle\int_0^{2\pi}\mathrm d\Phi\,
\mathcal K_1(\Phi)D_1^{\mathrm H}(P_s,\Phi)\right|^2
}{
N_h^2\left|\displaystyle\int_0^{2\pi}\mathrm d\Phi\,
\mathcal K_7(\Phi)\right|^2
}.
\label{eq:59}
\end{equation}
The harmonic-dependent single-electron responses and phases within a coil are both included in $\mathcal K_N$.

\section{Generality of harmonic selection in layered structures}

A source with a finite number of coils can be regarded as an axial phased array of repeated source elements. More generally, suppose that the $j$th identical source layer is translated by $z_j$ along the axis and rotated by $\delta_j$ about it. Sampling the source phase contributes $Nm_L\delta_j$, while the Fourier projection onto source mode $p$ contributes $-p\delta_j$. Let the normalized electron-number weights be $w_j\geq0$, with $\sum_jw_j=1$, and let $\rho_j(\vartheta)$ be the density of the additional injection phase $\vartheta$. Define $b_{N,j}=\int_0^{2\pi}\rho_j(\vartheta)e^{iN\vartheta}\mathrm d\vartheta$. Provided that all layers have identical single-electron trajectories, transverse envelopes, and detection projections, the layer factor is
\begin{equation}
F_{N,p}=\sum_{j=0}^{L-1}w_jb_{N,j}
\exp\left\{i\left[N\kappa z_j+(Nm_L-p)\delta_j\right]\right\}.
\label{eq:60}
\end{equation}
Relative azimuthal rotations make this factor a joint filter for harmonic order and azimuthal mode. In the absence of relative rotations between layers, $\delta_j=0$, the factor is independent of $p$, so longitudinal selection does not relabel the original azimuthal modes. For a slice of the helical source considered here at a fixed azimuth $\Phi$ within a coil, $z_j=jP$, $w_j(\Phi)=w_{\mathrm H}(2\pi j+\Phi)/N_h$, and $b_{N,j}=1$. The layer factor is therefore $D_N^{\mathrm H}(P,\Phi)/N_h$.

As a discrete example, consider equally spaced layers, $z_j=z_0+j\Delta z$, with $L\geq2$ and a common bunching factor $b_N$. If Hann weights are sampled at the layer centers, then
\begin{equation}
\begin{aligned}
w_j&=\frac{1-\cos[2\pi(j+1/2)/L]}{L},
\qquad a=N\kappa\Delta z,
\\
F_N&=\frac{b_Ne^{iN\kappa z_0}}{L}
\left[
D_L^{(0)}(a)
-\frac{1}{2}e^{i\pi/L}D_L^{(0)}\left(a+\frac{2\pi}{L}\right)
\right.
\\
&\hspace{3.5em}\left.
-\frac{1}{2}e^{-i\pi/L}D_L^{(0)}\left(a-\frac{2\pi}{L}\right)
\right].
\end{aligned}
\label{eq:61}
\end{equation}
When $L=MN_\star$ with integer $M\geq2$, $\kappa\Delta z=2\pi s/N_\star$, and $\gcd(|s|,N_\star)=1$, only orders that are multiples of $N_\star$ are retained, consistent with the weighted finite phase sum in Eq.~\eqref{eq:53}.

To control $b_{N,j}$ within a finite harmonic band $1\leq N\leq H$, a real, normalized model is
\begin{equation}
\rho_j(\vartheta)=\frac{1}{2\pi}
\left[
1+2\alpha_b\operatorname{Re}\left(
e^{i\psi_j}\sum_{N=1}^{H}e^{-iN\vartheta}
\right)
\right],
\qquad 2\alpha_bH\leq1.
\label{eq:62}
\end{equation}
The sufficient condition for positivity ensures $\rho_j\geq0$, and the in-band bunching factors are $b_{N,j}=\alpha_be^{i\psi_j}$.

\section{Vortex phase, vector angular momentum, modal power, and a finite driving pulse}

Intercoil matching fixes the source-phase slope within a coil, thereby linking harmonic selection to azimuthal phase. At the target direction and frequency, Eq.~\eqref{eq:52} gives
\begin{equation}
\begin{aligned}
p&=B_{N_\star}=N_\star m_L+s,
\\
\ell_{N_\star}&=p+(N_\star-1)\epsilon
=N_\star m_L+(N_\star-1)\epsilon+s.
\end{aligned}
\label{eq:63}
\end{equation}
Here $\ell$ is the scalar phase index of $e^{i\ell\phi}$ in the projection onto the fixed basis vector $\bm e_\epsilon$. For H7, $\ell_7=7m_L+6\epsilon+s$: the reference choice $(m_L,\epsilon,s)=(0,-1,-1)$ gives $\ell_7=-7$, while a scan of the positive pitch branches $s=1,\ldots,5$ gives $-5,-4,-3,-2,-1$. The effects of varying $m_L$ and the branch index appear separately in this expression.

For the full vector field, the spin projections of the polarization basis must also be included. In the fixed basis used here, $\bm e_\epsilon$ is an eigenvector of $S_z$ with eigenvalue $\epsilon$. The corresponding pure vector mode therefore has the total angular-momentum index, in units of $\hbar$,
\begin{equation}
j_z=\ell_{N_\star}+\epsilon
=p+N_\star\epsilon
=N_\star(m_L+\epsilon)+s.
\label{eq:64}
\end{equation}
The scalar exponent of the other transverse component is $p+(N_\star+1)\epsilon$, and that of the longitudinal component is $p+N_\star\epsilon$. After their respective spin projections are included, all three components have the same $j_z$. This is consistent with the angular-momentum addition structure in radiation from cold helical beams~\cite{bogdanov2020}.

For the actual finite-pulse field, use fixed Cartesian components $\alpha=x,y,z$ and define
\begin{equation}
\begin{aligned}
E_{\alpha,\ell}(k,u)
&=\frac{1}{2\pi}\int_0^{2\pi}\mathrm d\phi\,
\widetilde E_\alpha(k,u,\phi)e^{-i\ell\phi},
\\
\mathcal P_\ell
&=\sum_\alpha\int_{\mathcal B}\mathrm dk\,W(k)
\int_0^{u_{\max}}\sin u\,\mathrm du\,
|E_{\alpha,\ell}(k,u)|^2.
\end{aligned}
\label{eq:65}
\end{equation}
The full-vector OAM modal power fractions reported in the main text are then evaluated as $\eta_\ell=\mathcal P_\ell/\sum_{\ell'}\mathcal P_{\ell'}$. Here $\mathcal B$ is an explicitly selected frequency band, and $W(k)$ is the spectral weight.

In terms of the optical phase $\eta=k_L\chi$, the reference flat-top envelope is
\begin{equation}
F(\eta)=
\begin{cases}
1, & |\eta|\leq40\pi,\\
\dfrac{1}{2}\left[1+\cos\left(\dfrac{|\eta|-40\pi}{8}\right)\right],
& 40\pi<|\eta|<48\pi,\\
0, & |\eta|\geq48\pi.
\end{cases}
\label{eq:66}
\end{equation}
Within the slowly varying approximation in the focal region, the vector potential can be expressed by multiplying Eq.~\eqref{eq:22} by a Gaussian envelope and $F$. The electromagnetic fields follow from the spatial and temporal derivatives of the potential.

\phantomsection